\documentclass{aa}
\usepackage{graphicx}
\usepackage{epsfig}
\begin{document}
   \title{The high energy  emission line spectrum of NGC~1068}

   \author{Giorgio Matt
          \inst{1}
          \and
          Stefano Bianchi\inst{1}
	  \and
	  Matteo Guainazzi\inst{2}
	  \and
 	  Silvano Molendi\inst{3}
          }

   \offprints{G. Matt}

   \institute{Dipartimento di Fisica, Universit\`a degli Studi Roma Tre,
via della Vasca Navale 84, I-00146 Roma, Italy\\
         \and
      XMM-Newton Science Operation Center/RSSD-ESA, Villafranca del Castillo,
Spain \\
	\and
IASF--CNR, Via Bassini 15, I--20133, Milano, Italy
             }

   \date{Received ; accepted }

   \abstract{We present and discuss the high energy ($E>$4 keV) XMM--$Newton$ 
spectrum of the Seyfert 2 galaxy, NGC~1068. Possible  evidence for flux variability in both
the neutral and ionized reflectors with respect to a BeppoSAX observation taken 3.5 years before is
found. Several Fe and Ni emission lines, from both neutral and highly ionized material, are
detected. The intensity of the iron K$\alpha$ Compton shoulder implies that the 
neutral reflector is Compton--thick, likely the visible inner wall of the $N_H > 10^{25}$ cm$^{-2}$
absorber. From the equivalent width of the ionized iron lines a column density of a 
few$\times 10^{21}$ cm$^{-2}$ is deduced for the hot ionized reflector. Finally, an iron (nickel)
overabundance, when compared to solar values, of about 2 (4) with respect to lower Z
elements, is found.
   \keywords{galaxies: individual: NGC~1068 - galaxies: Seyfert - X-rays: galaxies}
               }

   \maketitle
%
%________________________________________________________________

\section{Introduction}

The archetypal Seyfert 2 galaxy, NGC~1068, has been extensively studied at
all wavelengths. It was the discovery of broad lines in the polarized flux of
this source that led Antonucci \& Miller (1985) to propose the Unification model
for Seyfert galaxies, which is at present the most popular scenario for this class
of sources. In X--rays, after the pioneering $Einstein$ and EXOSAT observations
(Monier \& Halpern 1987; Elvis \& Lawrence 1988),  the $GINGA$ discovery of
a strong iron line (Koyama et al. 1989), being interpreted as due to the reprocessing
by circumnuclear matter of the otherwise invisible nuclear radiation, brilliantly
confirmed the Antonucci \& Miller model. ASCA (Ueno et al. 1994; Iwasawa et al. 1997; 
Bianchi et al. 2001) demonstrated that the line is complex, implying reflection from 
both neutral  and ionized material. The two reflectors model was confirmed, observing the
hard X--ray continuum, by BeppoSAX (Matt et al. 1997a), which was also able to put a
lower limit of about 10$^{25}$ cm$^{-2}$ to the column density of the absorbing matter.
Guainazzi et al. (1999) and Bianchi et al. (2001)
demonstrated that the situation is even more complex, the line spectrum requiring at
least three reflectors, one neutral (cold), one mildly ionized (warm) and one
highly ionized (hot).
Comparing the two BeppoSAX observations, taken about one year apart, Guainazzi et al. (2000)
revealed a flux variability, which they interpreted as due to a variation of the intensity of
the hot ionized reflector, so placing an upper limit to its size of the order of a parsec
or so.

\begin{figure*}[t]
\begin{minipage}{90mm}
\epsfig{file=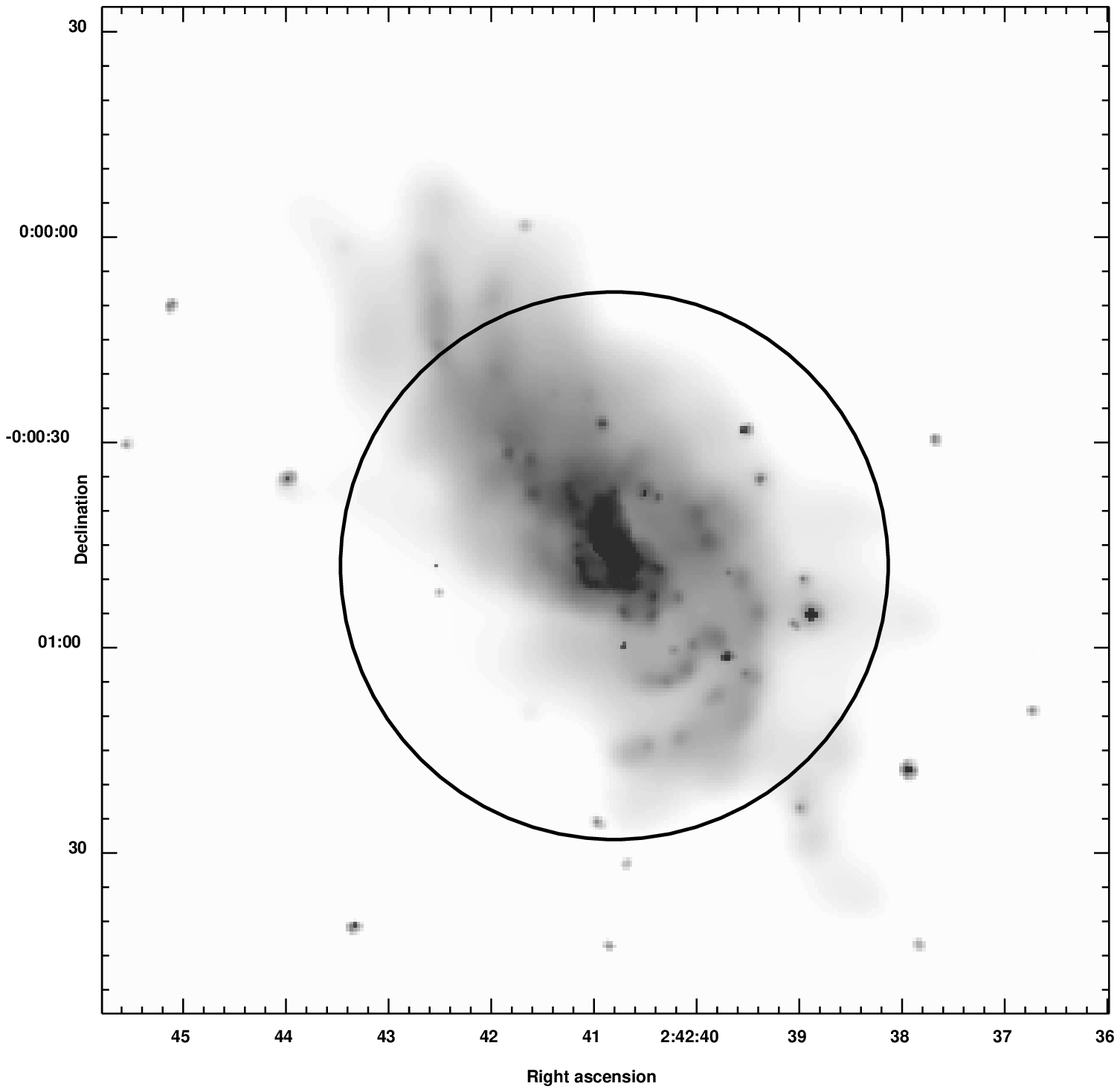,height=9.cm,width=9.4cm}
\end{minipage}
%\hfill
\begin{minipage}{90mm}
\epsfig{file=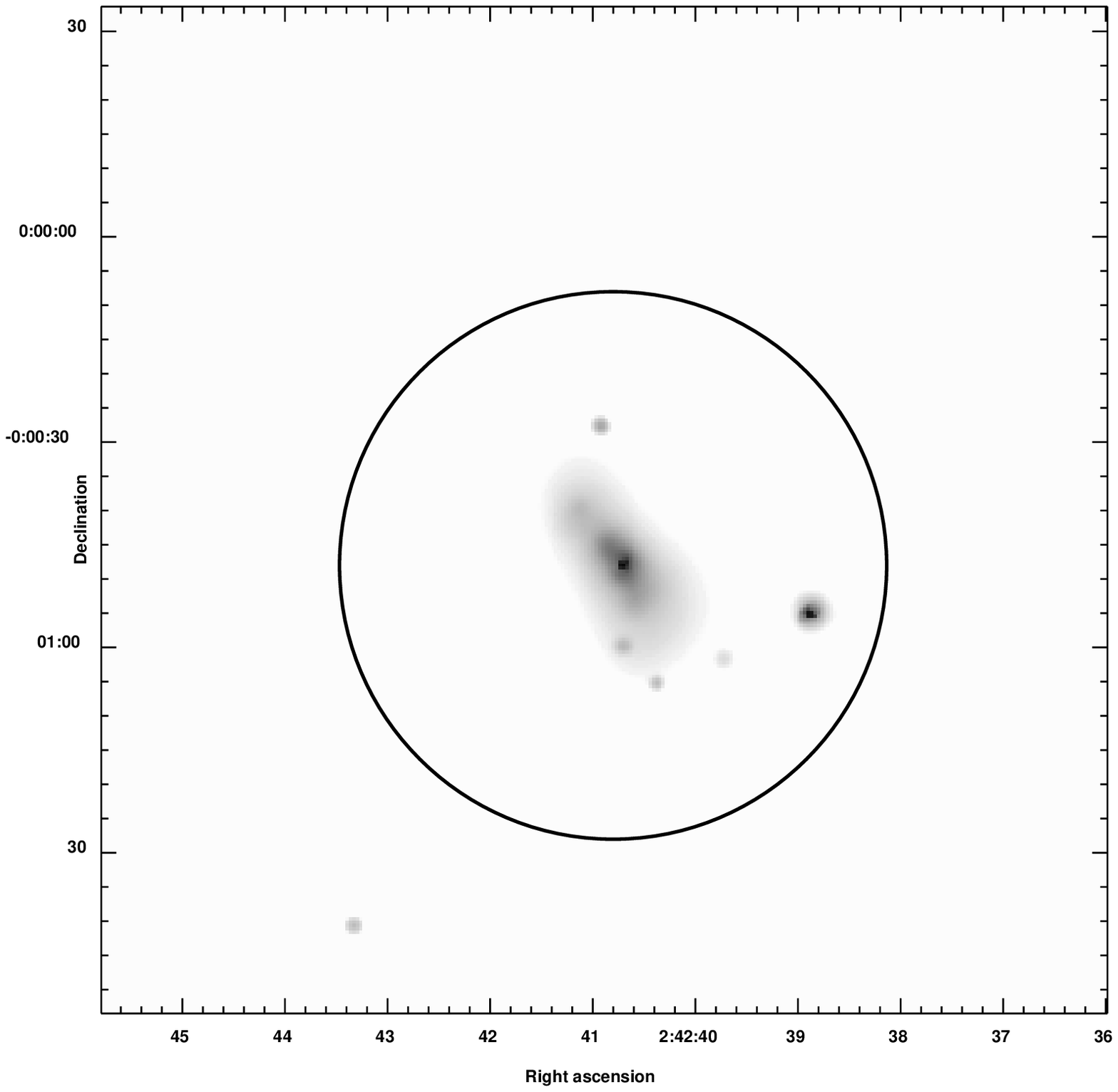,height=8.75cm,width=9.1cm}
\end{minipage}
\caption{{\it Left:} The whole band $Chandra$ ACIS image of the nuclear region of NGC~1068.
{\it Right:} The same, but only above 4 keV. Note that the extended emission is less
prominent, and that one off-center source, CXJ024238.9-000055, becomes by far the most
prominent one. }
\label{image_chandra}
\end{figure*}

High spatial and spectral resolution $Chandra$ and XMM--$Newton$ observations have 
confirmed the complexity of the circumnuclear region. The $Chandra$ image (Young et al. 2001)
revealed, especially in soft X--rays, a very rich morphology, with the brightest spot, 
however, confined in a 1\arcsec.5 (corresponding to 118 pc assuming $H_0$=70 km/s/Mpc, 
Young et al. 2001) region around the nucleus. Indeed, the iron emission as well as
the neutral and ionized continuum reflection are mostly 
confined in the nuclear region (Ogle et al.
2003). Many off--center point--like sources are also detected (Smith \& Wilson 2003). 

Gratings observations (Kinkhabwala et al. 2002; Ogle et al. 2003) have measured a line
spectrum which is consistent with photoionized plasma with a wide range of ionization
parameters. The line fluxes require significant contribution from resonant scattering
(Kinkhabwala et al. 2002), as predicted by Band et al. (1990) and Matt et al. (1996).
Given the limited energy range of the XMM--$Newton$/RGS and the modest effective area
of the $Chandra$/HETG at high energies, these results mainly concern the soft X--ray 
line spectrum. 

In this paper we analyse and discuss the high energy ($>$4 keV) XMM--$Newton$
spectrum of NGC~1068. While the CCDs of the EPIC cameras 
have a significantly poorer energy resolution than the
$Chandra$ HETG (the only grating instrument currently working at the
iron K$\alpha$ energy), for many purposes 
the much larger effective area above 4 keV overcompensate
for the lower resolution. As an example, one can compare the wealth of informations
on Fe and Ni K lines derived from the XMM--$Newton$ observation of the Circinus Galaxy
(Molendi et al. 2003) with
the much poorer informations on the same lines (indeed many of them not even detected) derived
from the $Chandra$/HETG spectrum (Sambruna et al. 2001).

\section{Data reduction}

\begin{figure*}[t]
\begin{minipage}{85mm}
\epsfig{file=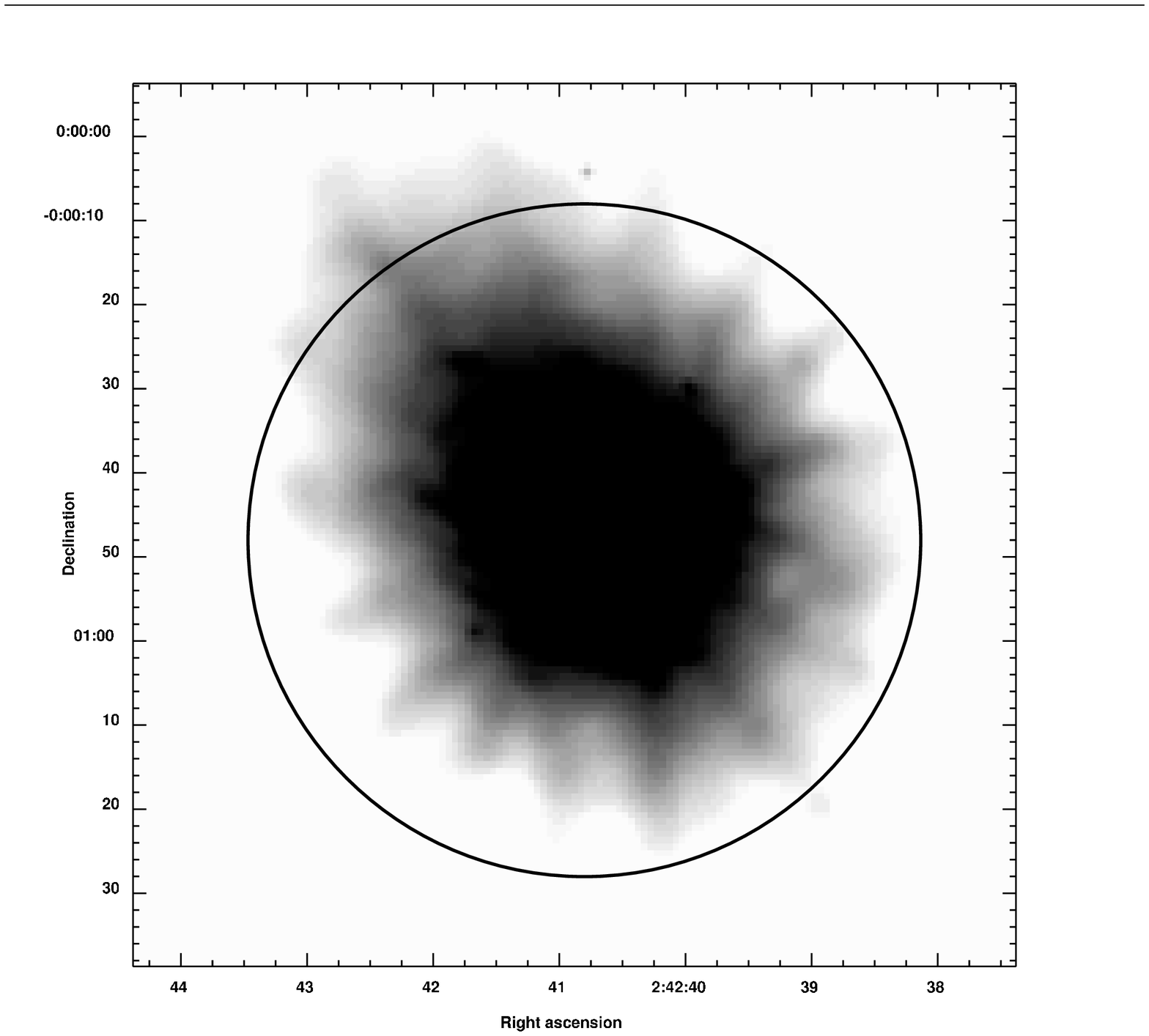,height=9.cm,width=9.4cm}
\end{minipage}
\begin{minipage}{85mm}
\epsfig{file=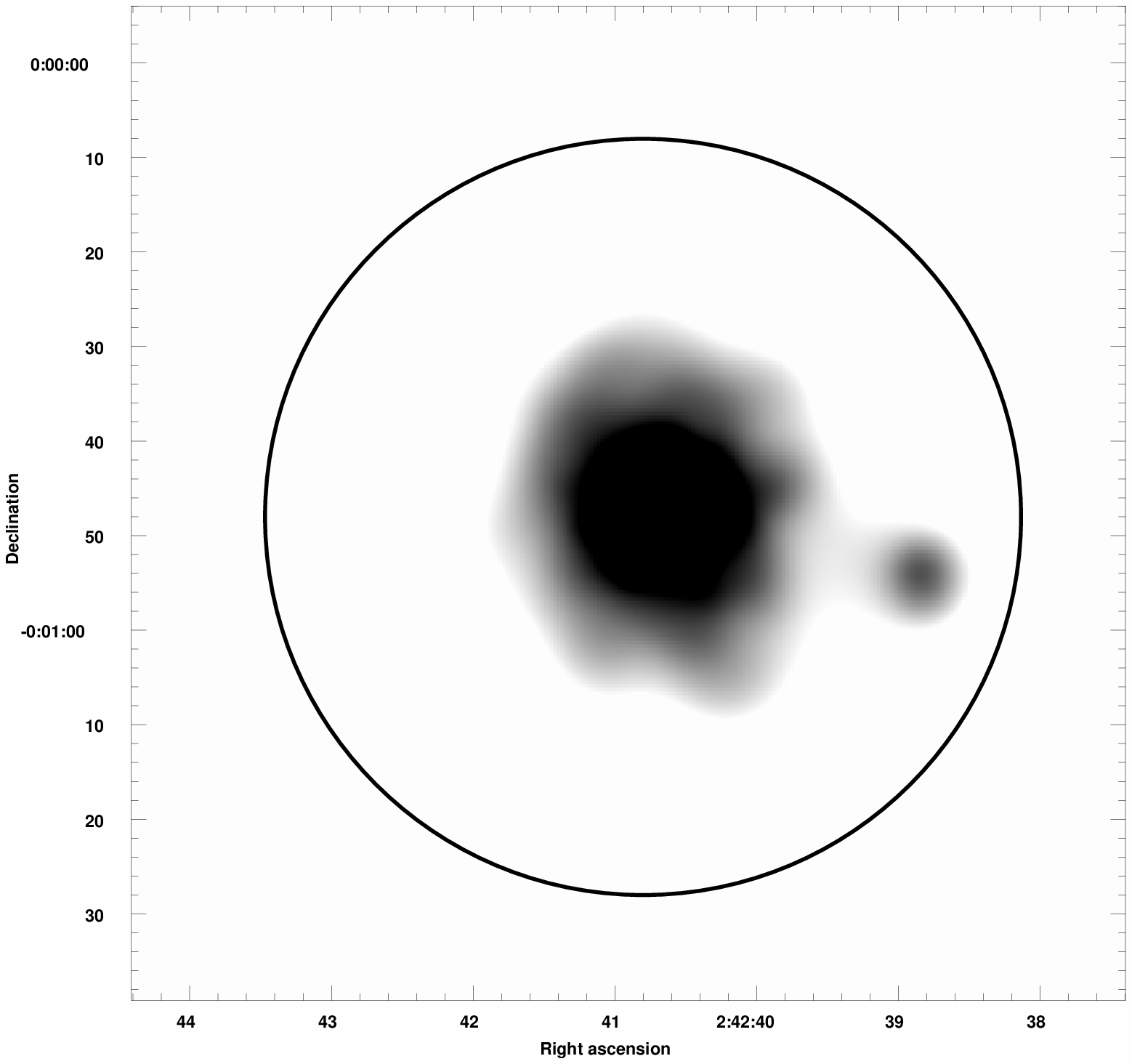,height=8.7cm,width=9.1cm}
\end{minipage}
\caption{{\it Left:} The whole band XMM--$Newton$ EPIC-pn image of the nuclear region of NGC~1068.
{\it Right:} The same, but only above 4 keV}.
\label{image_xmm}
\end{figure*}

\subsection{XMM--Newton}

NGC~1068 was observed twice by XMM-\textit{Newton} on July 29, 2000 and 
July 30, 2000 with the imaging CCD cameras, the EPIC-pn and MOS, adopting the Medium filter 
and operating in Large Window (pn), Full Window (MOS1) and Prime Partial W2
(MOS2). We retrieved the ODF files from the public archive, and reduced them
with the latest available SAS version (5.4.1), therefore using the most updated
calibrations.  The observed count rate is lower than the 
maximum defined for a 1\% pileup for the selected pn subarray (see Table 3 
of XMM-\textit{Newton} Users' Handbook), but it is close to that value, 
so we conservatively decided to exclude patterns higher than 0 in data 
reduction. An analysis with the \textsc{epatplot} tool indicates that
same residual pile--up effect may still be present, so we extracted a
spectrum removing the inner 6\arcsec ~circle region.  No significant differences
are found, however, in the continuum and in the lines alike, if the comparison is limited
to the energy range of interest here, i.e.  above 4 keV. We therefore concluded
that, for the pn, pile--up is not relevant for our purposes.
On the other hand, we shall not use MOS data because MOS1
has been used in Full Window mode and therefore strongly
suffers from pile--up problems, while MOS2
does not add much information, expecially because we will limit our analysis to $E>$4 keV.
Moreover, the pn appears to be the best calibrated detector at the energies
under consideration, as deduced from the analysis of the  
iron line complex in the spectrum of the Circinus Galaxy (Molendi et al. 2003). 

Since the two observations are nearly contiguous and 
no significant flux variations are found in their lightcurve, we 
verified that the two spectra were consistent with each other and 
decided to combine them. We used the \textsc{SAS}  tool 
\textsc{merge} to obtain a single event file and then extracted a 
spectrum from a region of $40\arcsec$ ~radii. The total exposure time, 
after rejecting time intervals of flaring particle background, is 61 ks. 
Spectra were analysed with \textsc{Xspec} 11.2.0. 

\subsection{Chandra}

Given the complexity of the morphology of the circumnuclear region of NGC~1068,
we also analysed the \textit{Chandra} observation of NGC~1068 
performed on February 21, 2000 with the Advanced CCD Imaging Spectrometer 
(ACIS-S), for a total exposure time of 48 ks. The 
default 3.2 s frame time, though resulting in the nuclear spectrum spoiled 
by pileup, does not affect our qualitative imaging analysis and the 
spectrum of the much fainter source CXJ024238.9-000055 (see below). Data were reduced 
with CIAO 2.3, using CALDB 2.21. Spectra were analysed with 
\textsc{Xspec} 11.2.0.

\section{Spatial analysis}

In Fig.~\ref{image_chandra} the $Chandra$ image over the whole working band (left panel)
and above 4 keV only (right panel) is shown. The detailed spatial analysis is discussed by
Young et al. (2001), while the X-ray source population by Smith \& Wilson (2003).
The XMM--$Newton$ image (whole band, left panel, and above 4 keV, right panel) is shown in
Fig.~\ref{image_xmm}. In all images, the $r$=40\arcsec ~circle over wich the XMM--$Newton$ spectrum
has been extracted is shown. 

For our own purposes, it is important to note that: 
a) the nucleus is partly extended over several arcseconds even
above 4 keV, the extended emission accounting for about a quarter of counts in this band.
Our 40\arcsec ~XMM--$Newton$ extraction regions encompasses this extended emission. b) 
At a distance of about 28\arcsec, a 
bright point--like source (CXJ024238.9-000055) is clearly apparent
in the $Chandra$ image (hereinafter we refer to it for simplicity as X--1); in the hard band,
it is by far the brightest off--nuclear point--like  source.
Smith \& Wilson (2003) derived for this source
a 0.4-8 keV flux of about 5$\times10^{-13}$ erg cm$^{-2}$ s$^{-1}$, corresponding
to a luminosity, at the distance of NGC~1068, of 1.4$\times10^{40}$ erg s$^{-1}$. The spectrum
(which we reanalysed) is quite hard, being 
consistent with either a very flat ($\Gamma$=0.85) power law or
with  multicolor disc blackbody with an inner temperature of about 4 keV. While we defer
the reader to Smith \& Wilson (2003) for a discussion of its possible nature,
we note here that this source, being so hard, may significantly contaminate the spectrum
of the NGC~1068 nucleus in the energy range under consideration in this paper. 
Indeed, while the source is not readily visible in the whole band XMM--$Newton$ image, 
it is clearly present
in the $E>$4 keV image. To evaluate its contribution to the overall spectrum,
we extracted a 20\arcsec ~spectrum centered on NGC~1068 and a 10\arcsec 
~spectrum centered on X--1. 
The flux we derived for NGC~1068 is 14 times larger than that of X--1 (which in turn is about half
that measured by $Chandra$, Smith \& Wilson 2003), and this is
certainly a lower limit as the spectrum of X--1 is heavily contaminated by that of NGC~1068.
 In fact, in the X--1 spectrum a broad iron line is clearly present, well fitted
by two narrow components, one at 6.4 keV (neutral iron) and one at 6.7 keV
(He--like iron).  If we assume that they are the result of the contamination
by NGC~1068 (no iron lines are apparent in the, admittedly poor, $Chandra$ spectrum),
this can provide a method to estimate the level of contamination of X-1 by the nucleus. 
The EWs for these two lines
are about 2.5 smaller than in the NGC~1068 spectrum (when calculated with respect to the
whole continuum; see next section). 
Therefore, about 40\% of the spectrum extracted around X--1 is actually due to
NGC~1068, which implies that X-1 does not contaminate the 40\arcsec 
~spectrum of NGC~1068 by more than
about 5\%. Considering also that we are mostly interested in the emission lines, which are
unlikely to be produced by X--1, we decided that this contamination is not large enough to
justify the 20\% loss in photons when using the $r=20\arcsec$ ~region.
In the following we will then use the 40\arcsec ~radius spectrum, unless explicitely stated. 

\section{The E$>$4 keV XMM--$Newton$ spectrum}

The soft X--ray high energy resolution line spectrum of NGC~1068 is discussed in detail
by Kinkhabwala et al. (2002) and Ogle et al. (2003).
We limit ourselves to analyse and discuss the high energy  ($E>$4 keV), moderate
resolution EPIC--pn spectrum of
the nuclear region of NGC~1068, with particular emphasis on the iron and nickel
line properties. Let us, however, first briefly discuss the continuum properties.

\subsection{The continuum}

\begin{table}[t]
\caption{Best fit results. Continua} 
\begin{tabular}{||l|c||}
\hline
& \cr
$\Gamma$ & 2.04$^{+0.04}_{-0.03}$  \cr
& \cr
$N_{\rm cold}$ (1 keV) [ph keV$^{-1}$ cm$^{-2}$ s$^{-1}$] & 0.019 \cr
& \cr
$N_{\rm ion}$ (1 keV) [ph keV$^{-1}$ cm$^{-2}$ s$^{-1}$] & 7.37$10^{-4}$ \cr
& \cr
$F_{\rm cold}$ (4--10 keV)  [erg cm$^{-2}$ s$^{-1}$] & 1.52$\times10^{-12}$ \cr
& \cr
$F_{\rm ion}$ (4--10 keV)  [erg cm$^{-2}$ s$^{-1}$] & 1.01$\times10^{-12}$ \cr
& \cr
A$_{\rm Fe}^{a}$ & 2.39$^{+0.43}_{-0.36}$ \cr
& \cr
\hline
\end{tabular}
~\par
$^{a}$ in solar units (Anders \& Grevesse 1989) by number.
\label{bestfit_c}
\end{table}

\begin{figure}
\epsfig{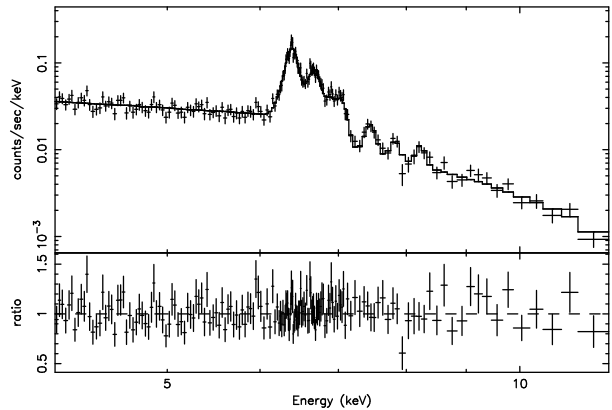}
\caption{Spectrum and best fit model, and model/data ratio. See text for details.}
\label{spec}
\end{figure}

\begin{figure}
\epsfig{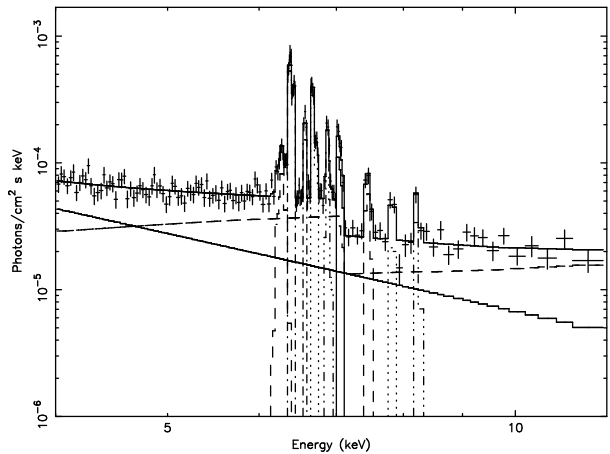}
\caption{Unfolded spectrum and best fit model. See text for details.}
\label{ufold}
\end{figure}

Following Iwasawa et al. (1997) and Matt et al. (1997a), we fitted the continuum with two
components, describing reflection from both neutral and ionized matter (the latter component
including both the `warm' and `hot' regions, Bianchi et al. 2001). For 
the neutral reflection component we adopted, as customary, the model {\sc pexrav}, with $R$
fixed to --1 (i.e. with the primary emission shut off). The
ionized reflector component is instead modeled by a simple power law, as the ionized reflector
should act as a ``mirror'' for the primary emission. The power law index has been assumed
to be the same for the two models. To fit the spectrum, we of course added as many lines
as required, as discussed in details in the next paragraphs. The fit is fully acceptable
($\chi^2$/d.o.f.=143.7/149). The best fit parameters for the continua, for the neutral
and the ionized lines are reported in Tables~\ref{bestfit_c}, \ref{bestfit_lcold} and
\ref{bestfit_lion}, respectively (all errors refer to 90\% confidence level for one
interesting parameter, unless explicitely stated). The folded spectrum and best fit model, along
with the model/data ratio, are shown
in Fig.~\ref{spec}, while the unfolded spectrum is shown in Fig.~\ref{ufold}.
  The power law index is 2.04$^{+0.04}_{-0.03}$, and the ratio
between the 4--10 keV cold and ionized reflectors fluxes is about 1.5. (Fitting the
spectrum extracted from a 20\arcsec ~radius, a slightly steeper index, 2.10$^{+0.04}_{-0.04}$, is found;
the ratio between the two components does not change significantly. 
All other parameters are also consistent with those obtained with the 
$r$=40\arcsec ~extraction radius, but with
of course somewhat larger error bars). This result could be compared
with that derived by Matt et al. (1997a) from the December 1996
BeppoSAX observation, where the flux ratio was instead 0.5,
suggesting significant variations in likely both components. Indeed, variability on a time
scale of 1 year, probably in the ionized reflector component, has been reported by Guainazzi
et al. (2000). However, the BeppoSAX result must be taken with caution: the 
$F_{\rm cold}$/$F_{\rm ion}$
ratio is very sensitive to the power law index, which in BeppoSAX was only loosely
determined (1.74$^{+0.25}_{-0.56}$). We re--fitted the BeppoSAX spectrum imposing the power
law index found from XMM--$Newton$, i.e. 2.04. The result is only slightly worse statistically
$\Delta\chi^2$=2.7 with one less free parameter, corresponding to a 90\% confidence
level according to the F-test),
and the ratio now becomes $\sim$1.3. On the contrary, fitting the XMM--$Newton$ spectrum imposing
the BeppoSAX best fit power law index gives a significantly
worst fit ($\chi^2$/d.o.f.=174.4/150); the ratio is now about 1.15. Indeed, 
the XMM--$Newton$ 90\% confidence level lower limit on $\Gamma$ (i.e. 2.01)
corresponds to a flux ratio of 1.49.  
At this confidence level, therefore, the BeppoSAX and XMM--$Newton$ estimates are 
inconsistent with each  other.  There is therefore some, even if still not conclusive given also the
uncertainties when comparing different instruments, evidence 
for variability between the BeppoSAX and the XMM--$Newton$ observations, taken 3.5 years apart,
which would give an upper limit to the distance of the spatially unresolved
reflecting regions of the order of a few parsecs. Clearly, a second XMM--$Newton$ observation
is necessary to confirm this finding.

\begin{figure}
\epsfig{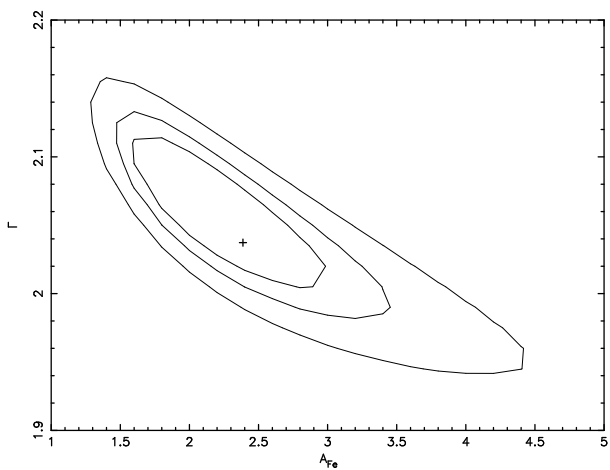}
\caption{$A_{\rm Fe}$--$\Gamma$ contour plot.}
\label{afe_gamma}
\end{figure}

%\begin{figure}
%\epsfig{file=afe_i.ps,height=8.cm,angle=-90}
%\caption{$A_{\rm Fe}$--$i$ contour plot. }
%\label{afe_i}
%\end{figure}

Another interesting result is the large iron abundance, $A_{\rm Fe}$=2.4 (with respect to
the solar value of Anders \& Grevesse 1989). In the fit, $A_{\rm Fe}$ is basically
derived from the 
depth of the iron edge in the Compton reflection component, which depends on the spectral
index and, for a plane--parallel geometry, also on the
inclination angle. 
Solar abundances for the other elements (in particular for CNO elements, which
contribute most to the photoelectric cross section at 6.4 keV) have been assumed. 
The fit has been performed
keeping the inclination angle $i$ fixed to 63$^{\circ}$, i.e. the {\sc Xspec} default value. For
lower values of the inclination angle a less deep edge is expected, and therefore an even
larger value of the iron abundance is found (e.g. 3.3 for an inclination of about 20$^{\circ}$).
In Figs.~\ref{afe_gamma} the $A_{\rm Fe}$--$\Gamma$ (with $i$ free to vary) is shown; it is
important to note that the iron abundance is larger than 1.5 at the 99\% confidence level (two 
interesting parameters).

\subsection{Emission lines from the cold reflector}

Of the several lines observed in the high energy spectrum of NGC~1068 (see Fig.~\ref{spec}), four
can be attributed to neutral Fe and Ni atoms, i.e. the K$\alpha$ and K$\beta$ lines from
both elements (see Table~\ref{bestfit_lcold}). 
All lines, both neutral and ionized, have been fitted with a gaussian profile
with $\sigma$ fixed to 1 eV (i.e. much less than the energy resolution;
no improvement is found leaving $\sigma$ free to vary for the brigthest lines). 
All lines we attribute to the neutral reflector are significant at more than 99.99\%
confidence level, according to the F-test, apart from the Ni K$\beta$ at 8.25 keV which is
significant at the 99.90\% confidence level.

The iron K$\alpha$ line energy (6.424$\pm$0.001 keV; note that the statistical error
is much lower than the calibration uncertainty, which is of the order of 10 eV\footnote{see
{\tt http://xmm.vilspa.esa.es/docs/\-documents/\-CAL-TN-0018-2-1.pdf}.}) is larger than the
value for neutral iron (actually a doublet with energies of 6.391 and 6.405 keV, House 1969,
with a 1:2 branching ratio), and would instead suggest Fe {\sc xvi}; however, 
the measured K$\beta$/K$\alpha$ flux ratio for iron (including only the line
core for the K$\alpha$ line, see below) is 0.20$^{+0.07}_{-0.03}$,
slightly larger than the expected value even for neutral value, which is
about 0.16 (see Molendi et al. 2003). This value stays almost constant up
to Fe {\sc ix}, then starts decreasing (Kaastra \& Mewe 1993). It is therefore
unlikely that iron is much more ionized than that (the line energy is also basically
constant over this ionization range, House 1969). The discrepancy with the observed
line energy may result from a less than perfect energy calibration of the pn,
or a real blueshift of the line, which is however hard to explain in the classical torus
model for the neutral reflection. It is worth noting that the MOS2
gives exactly the same best fit line energy, and
that $Chandra$/HETG also measured a blueshift, albeit smaller (6.411$\pm$0.007 keV,
Ogle et al. 2003). The K$\beta$/K$\alpha$ 
ratio for nickel is 0.57$\pm$0.35, definitely larger than the expected value, which is similar
to that for iron (Kaastra \& Mewe 1993). While the measured energy of the line, 
8.25$^{+0.10}_{-0.03}$ keV, 
is perfectly consistent with the expected value of 8.265 keV for the neutral Ni
K$\beta$ line (Bearden 1967), the Fe {\sc xxvi} K$\beta$ lines at energies clustering around
8.25
\footnote{Unless explicitely stated, line energies are taken from 
the NIST Atomic Spectra Database,
{\tt http://physics.nist.gov/cgi-bin/AtData/main\_asd.}} are also possibly contributing. 

The Ni--to--Fe  K$\alpha$ flux ratio is 0.13$\pm$0.03, much larger than the expected value of
0.04-0.06 (calculated as described in Molendi et al. 2003 assuming a factor 2 overabundance
for both iron and nickel). This implies a nickel--to--iron overabundance of a factor $\sim$2,
or more

Clear evidence for an iron K$\alpha$ Compton shoulder (hereinafter
CS) is also found, confirming and refining the ASCA (Iwasawa et al. 1997) and $Chandra$/HETG
(Ogle et al. 2003) results. 
 The CS is fitted with a gaussian profile with the centroid energy and $\sigma$
fixed to 6.3 keV and 40 eV, respectively (Matt 2002).
If the CS is not included 
a significantly worse fit ($\chi^2$/d.o.f.=181.2/149) is found, even allowing for
the width of the main line to vary. The ratio between the CS and the line core fluxes is about
0.2, in agreement with the theoretical expectation for Compton--thick reflecting material
(Matt 2002), so suggesting an origin in the visible part of the $N_H >$10$^{25}$ cm$^{-2}$
absorbing material (Matt et al. 1997a). 

The iron K$\alpha$ EW (with respect to the cold reflection continuum)
is 1.2 keV (1.45 keV including the CS), reasonably in agreement with theoretical
expectations (e.g. Ghisellini et al. 1994, Matt et al. 2003) even with the
observed iron overabundance, given the less than linear
increase of the line EW with the iron abundance (Matt et al. 1997b).

\begin{table}[t]
\caption{Best fit results. Lines from cold matter. Equivalent widths are
calculated against the cold reflector only.} 
\begin{tabular}{||l|c||}
\hline
& \cr
E (Fe K$\alpha$ core) [keV] & 6.424$^{+0.001}_{-0.001}$ \cr
& \cr
F (Fe K$\alpha$ core)  [10$^{-6}$ ph cm$^{-2}$ s$^{-1}$] & 44.3$^{+2.3}_{-3.0}$ \cr
& \cr
EW (Fe K$\alpha$ core) [eV] & 1200 \cr
& \cr
F (Fe K$\alpha$ CS) [10$^{-5}$ ph cm$^{-2}$ s$^{-1}$] & 8.7$^{+1.3}_{-2.7}$ \cr
& \cr
E (Fe K$\beta$) [keV] & 7.077$^{+0.003}_{-0.045}$ \cr
& \cr
F (Fe K$\beta$) [10$^{-6}$ ph cm$^{-2}$ s$^{-1}$] & 9.1$^{+2.1}_{-1.1}$ \cr
& \cr
EW (Fe K$\beta$) [eV] & 346 \cr
& \cr
E (Ni K$\alpha$) [keV] & 7.48$^{+0.01}_{-0.05}$ \cr
& \cr
F (Ni K$\alpha$) [10$^{-5}$ ph cm$^{-2}$ s$^{-1}$] & 5.6$^{+1.8}_{-1.0}$ \cr
& \cr
EW (Ni K$\alpha$) [eV] & 410 \cr
& \cr
E (Ni K$\beta$) [keV] & 8.25$^{+0.10}_{-0.03}$ \cr
& \cr
F (Ni K$\beta$) [10$^{-5}$ ph cm$^{-2}$ s$^{-1}$] & 3.2$^{+1.1}_{-1.5}$ \cr
& \cr
EW (Ni K$\beta$) [eV] & 230 \cr
& \cr
\hline
\end{tabular}
\label{bestfit_lcold}
\end{table}

\subsection{Emission lines from the ionized reflector}

Four emission lines from highly ionized matter (the `hot' reflector cited above)
are also detected in the pn spectrum (Table~\ref{bestfit_lion}). All 
of them are significant at more than 99.99\%
confidence level, according to the F-test, apart from the Ni He--like K$\alpha$ at 7.83 keV which 
is significant at the 99.79\% confidence level. Besides
the He-- and H--like Fe K$\alpha$ lines\footnote{Also for these lines 
the measured energies are not in agreement with the expected values, if only
statistical errors are taken into account. However, in these cases identification
is certain, as no other strong lines are present at those energies. It is
worth noting that the energy of the He--like line is greater than expected, and that
of the H--like line lower, ruling out inflow or outflow if the two lines, as likely,
are produced in the same region.}, already discovered by ASCA (Ueno et al. 1994;
Iwasawa et al. 1997; Bianchi et al. 2001),
two lines at about 6.61 and 7.83 keV are also detected with high significance. The energy
of the first line points towards fluorescence from Be--like iron, which however has a quite
low fluorescent yield of 0.11 (Krolik \& Kallman 1987). Moreover, the matter may likely be
optically thick to the line, which results in an effective Auger destruction. It is therefore
possible that this line is actually a blend of Be-- and Li--like lines (the latter having
an energy of about 6.65, a high fluorescent yield, 0.75, and no possibility of Auger
destruction). (The possibility that this line is the CS of the He--like iron line is
ruled out by the low optical depth of the hot reflecting matter, as discussed below). 
The line at 7.83 keV is likely a blend of the Ni He--like K$\alpha$ 
(7.81 keV) and Fe He--like K$\beta$ lines (7.87 keV). The rather wide range of
ionization argues against collisional ionization. On the contrary, we have verified with
{\sc cloudy}\footnote{http://www.nublado.org/} that photoionization may
produce such a range in a single zone with e.g. a column density a few times 10$^{21}$ cm$^{-2}$
(see below) and a ionization parameter of about 150.

The measured values of the EWs can be compared with the results of the best previous observation,
i.e. the 100 ks ASCA observation (Bianchi et al. 2001). The results reported there are
somewhat different, but were obtained with a simpler model. We refitted the ASCA
spectrum with all the lines found by XMM--$Newton$ (and the power law index fixed
to 2.04) and found that, within the errors, the fluxes measured by ASCA are consistent
with those measured by XMM--$Newton$. It should also be noted that the presence of the 
6.61 keV line vindicates the claim of Iwasawa et al. (1997), based on the first ASCA
observation, of a redshift of the ionized lines. 

The EW of the He-- and H--like iron lines can be compared with the results of Matt et al.
(1996) to derive at least an order--of--magnitude value
for the column density of the reflecting material. 
In their figures 5 and 6 the EW of these lines, calculated 
including contributions from both recombination and
resonant scattering, are shown as a function of the column density of the reflecting
material. The calculations assumed iron abundance and fractions equal to 1 of the
considered ions. The measured iron overabundance roughly compensates
for the less than one fraction of each ion (for instance, our {\sc cloudy} calculation  
gives a value of 0.4 for the fraction of He--like iron atoms, and somewhat smaller value for 
the H--like atoms).  Even taking into account the fact that the 
ionized continuum includes also reflection from the warm reflector (Bianchi et al. 2001
estimated that the hot reflector contributes to about 3/4 of the total ionized reflection),
from the abovementioned figures of Matt et al. (1996) 
a column density of the hot ionized reflector of a 
few$\times10^{21}$ cm$^{-2}$ can be deduced.

\begin{table}[t]
\caption{Best fit results. Lines from ionized matter. Equivalent widths are 
calculated against the ionized reflector only.} 
\begin{tabular}{||l|c||}
\hline
& \cr
E (Fe Be-like K$\alpha$ ) [keV] & 6.61$^{+0.01}_{-0.04}$ \cr
& \cr
F (Fe Be-like K$\alpha$) [10$^{-5}$ ph cm$^{-2}$ s$^{-1}$] & 7.6$^{+2.4}_{-1.1}$ \cr
& \cr
EW (Fe Be--like K$\alpha$) [eV] & 485 \cr
& \cr
E (Fe He-like K$\alpha$ ) [keV] & 6.725$^{+0.001}_{-0.001}$ \cr
& \cr
F (Fe He-like K$\alpha$ ) [10$^{-5}$ ph cm$^{-2}$ s$^{-1}$] & 21.8$^{+3.2}_{-1.4}$ \cr
& \cr
EW (Fe He--like K$\alpha$) [eV] & 1430 \cr
& \cr
E (Fe H--like K$\alpha$) [keV] & 6.92$^{+0.01}_{-0.04}$ \cr
& \cr
F (Fe H--like K$\alpha$) [10$^{-5}$ ph cm$^{-2}$ s$^{-1}$] & 7.1$^{+2.8}_{-0.9}$ \cr
& \cr
EW (Fe H--like K$\alpha$) [eV] & 494 \cr
& \cr
E (Ni He-like K$\alpha$) [keV] & 7.83$^{+0.04}_{-0.05}$ \cr
& \cr
F (Ni He-like K$\alpha$) [10$^{-5}$ ph cm$^{-2}$ s$^{-1}$] & 2.7$^{+1.2}_{-1.3}$ \cr
& \cr
EW (Ni He--like K$\alpha$) [eV] & 246 \cr
& \cr
\hline
\end{tabular}
\label{bestfit_lion}
\end{table}

\section{Summary}

We have analysed the high energy ($>$4 keV) XMM--$Newton$ EPIC--pn spectrum of NGC~1068. 
The main results can be summarized as follows:

i) Possible (but not conclusive)
evidence for variations of both neutral and ionized reflectors with respect
to the December 1996 BeppoSAX observation is found, suggesting an upper limit to
the size of both reflectors of a few parsecs. Another XMM--$Newton$ observation, to compare
results obtained with similar instruments, is however required to confirm this finding. It is
worth noting that a variability on a time scale of about a year of the ionized reflector was
found by Guainazzi et al. (2000) comparing two BeppoSAX observations.

ii) Iron is overabundant, with respect to lower Z elements and when compared to the solar
value, by a factor about 2. Nickel is,
in its turn, overabundant by a factor $\sim$2 with respect to iron. A qualitatively 
similar result was found in the Circinus Galaxy (Molendi et al. 2003), where however both
overabundances were smaller.

iii) The iron K$\beta$/K$\alpha$ flux ratio points to a low ionization state of iron,
inconsistent with the K$\alpha$ line energy, suggesting calibration problems. The
nickel  K$\beta$/K$\alpha$ is larger than expected, but the K$\beta$ line may actually be a blend
with ionized iron lines. 

iv) The Fe K$\alpha$ Compton Shoulder is also detected, with a relative flux of 0.2,
in agreement with the value expected for reflection from Compton--thick matter (Matt 2002)

v) Be--, Li--, He-- and H--like iron emission lines, as well as He--like Ni line, 
are also found. Their EW suggests a column density of the hot ionized reflector of a 
few$\times10^{21}$ cm$^{-2}$ (Matt et al. 1996).

\section*{Acknowledgements}
Based on observations obtained with XMM--$Newton$, an ESA science mission with 
instruments and contributions directly funded by ESA Member States and the USA
(NASA). GM and SB acknowledge financial support from ASI.

\end{document}